\documentclass{ws-ijmpb}
\usepackage{bm}
\newcommand{\eqbreak}{
\end{multicols}
\widetext
\noindent
\rule{.48\linewidth}{.1mm}\rule{.1mm}{.1cm}
}
\newcommand{\eqresume}{
\noindent
\rule{.52\linewidth}{.0mm}\rule[-.1cm]{.1mm}{.1cm}\rule{.48\linewidth}{.1mm}
\begin{multicols}{2}
\narrowtext
}
\newcommand{\curD}{{\cal D}}

\newcommand{\curL}{{\cal L}}

\newcommand{\tc}{{T_{\rm c}}}

\newcommand{\phase}{\varphi}

\newcommand{\BdG}{Bogoliubov--de~Gennes}
\begin{document}
%-------------------------
\markboth{
Daniel E.~Sheehy}
{Feynman path-integral approach to the QED$_3$ theory of the pseudogap}

\title{Feynman path-integral approach to the QED$_3$ theory of the pseudogap}
\author{Daniel E. Sheehy}
\address{Department of Physics and Astronomy, 
University of British Columbia, 
6224 Agricultural Road, Vancouver, B.C.~V6T1Z1, Canada\\
sheehy@physics.ubc.ca}

\maketitle

\begin{history}
\received{(20 March 2003)}
%\revised{(Day Month Year)}
%\accepted{(Day Month Year)}
%\comby{(xxxxxxxxxx)}
\end{history}

%--------------------------
\begin{abstract}
%--------------------------
%--------------------------
 In this work the connection between vortex condensation in a d-wave superconductor
and the QED$_3$ gauge theory of the pseudogap is elucidated.  The approach
taken circumvents the use of the standard Franz-Tesanovic gauge transformation, borrowing 
ideas from the path-integral analysis of the Aharonov-Bohm problem.  
An essential feature of this approach is that gauge-transformations which are prohibited on
a particular multiply-connected manifold (e.g.~a superconductor with vortices) can be successfully 
performed on the universal covering space associated with that manifold.
%--------------------------
\end{abstract}
%--------------------------

\section{Introduction\/}
\label{SEC:intro}
%---------------------------
Recently, much attention has been focused on understanding 
the pseudogap~\cite{AGL96,HD96,CROF98,REF:Corson,REF:Xu} phenomena
 of the high-temperature superconductors.  One
possible  explanation~\cite{EK95,ML96,FM98,KD98} invokes
the notion that the pseudogap is due to the presence of pairing correlations
above the superconducting transition temperature $\tc$. 
 Within such a scenario, 
the lack of long-range phase coherence in the pseudogap is presumed to be due
to the proliferation of vortex excitations in it.    
Following the approach of many recent papers~\cite{FM98,KD98,REF:Balents,REF:Franz,REF:Herbut}, 
we consider the problem of coupling vortices to the quasiparticles of a d-wave
superconductor.  
As discussed in Refs.~\refcite{REF:Balents,REF:Franz}, 
there is an important distinction between the 
condensation of $hc/2e$ (i.e.~singly quantized) 
and $hc/e$ (i.e.~doubly quantized) vortices in a d-wave superconductor.  
Experimentally~\cite{REF:Bonn}, magnetic field-induced vortices seem to 
be exclusively of the singly quantized variety.  
We shall take this as evidence that, in
considering vortex excitations in the pseudogap regime, it is sufficient to consider
only $hc/2e$ vortices.

The technical issues associated with $hc/2e$ vortices have been discussed in 
Refs.~\refcite{REF:Balents,REF:Franz,REF:Franz2} and amount to the fact that certain singular
gauge transformations (which arise naturally when considering 
the \BdG\ equation in the presence of vortices) lead to the presence of 
quasiparticle branch cuts.
%--------------------------
Here we shall adopt the point of view that 
these difficulties are not unique to the problem of performing gauge transformations 
in the presence of vortices:  They arise when one considers the general problem
of making gauge-transformations on multiply-connected manifolds.
%--------------------------
For example, consider the Aharonov-Bohm problem~\cite{REF:AB}, in which one imagines
attaching current leads to a doubly-connected metallic ring through which a solenoid penetrates. 
Even in the case of a solenoid with a radius sufficently small that the magnetic field is negligible in the
ring, the presence of a nonzero vector potential leads to quantum-mechanical interference of electrons taking
different paths around the ring while  propagating between the leads.
Thus,
one cannot make a gauge-transformation which eliminates the effect of the vector potential on topologically
distinct Feynman paths.

In a similar sense, the winding of the pair-potential phase around a superconducting
vortex may not be trivially removed by a gauge transformation.  In particular, as noted above,
making certain singular gauge-transformations in the presence of $hc/2e$ vortices leads to 
quasiparticle branch cuts.
Our aim here is to discuss a new approach to handling such branch cuts using the topological properties of 
path integrals.
Recently, Franz and Tesanovic~\cite{REF:Franz2} have introduced a singular gauge transformation 
which avoids the introduction of quasiparticle branch cuts
via a clever trick involving splitting the vortices into two distinct groups ($A$  and $B$)
and transforming the electrons relative to group $A$ and the holes relative to group $B$.  These authors find
that the action governing the quasiparticle dynamics of a d-wave superconductor in the presence of fluctuating vortices 
is given by the well-known problem of 
three-dimensional quantum electrodynamics (QED$_3$): Dirac fermions coupled to a fluctuating 
\lq\lq Berry\rq\rq\ gauge
field $a$.  
However, as we shall discuss, the propagator is {\it not \/} given by the usual propagator for QED$_3$; it is 
given by the associated gauge-invariant propagator.  This occurs because $a$
is not a physical (i.e.~electromagnetic) gauge-field; thus the propagator must not transform under
gauge transformations on $a$.  Gauge-invariant propagators have appeared many times in the context of 
strongly-correlated electron 
systems~\cite{REF:Altshuler,REF:Rantner,REF:Khveshchenko}; in particular the gauge-invariant propagator for QED$_3$ 
exhibits intriguing non-Fermi liquid behavior~\cite{REF:Franz,REF:Khveshchenko}.

To obtain further insight into the physics of fluctuating vortices in d-wave superconductors, 
here we shall take a novel approach inspired by Schulman's~\cite{REF:Schulman} topological 
approach to the Aharonov-Bohm problem.
Our task is to shed light on the connection between fluctuating vortices and fluctuating gauge-fields.
Towards this end, we shall view a vortex as a \lq\lq hole\rq\rq\ which divides a superconductor into a 
doubly-connected space.  As in the Aharonov-Bohm problem, the fact that the superconductor is
multiply connected means that Feynman paths contributing to the quasiparticle 
propagator fall into topologically distinct sectors.  By making different gauge transformations
for topologically distinct Feynman paths, we shall see how the quasiparticle branch cuts 
may be expressed in terms of the Berry gauge field $a$. 
The purpose of this Paper is to arrive at the QED gauge theory of vortices in 
the pseudogap regime of the cuprates via a path integral technique.

This Paper is organized as follows: 
In Sec.~\ref{sec:bdg} we consider the problem of solving the Bogoliubov-de Gennes (BdG)
equation in the presence of a static array of vortices.  Although our main interest will be in the problem of 
vortex {\it fluctuations\/} in a d-wave superconductor, the static case will be sufficient to motivate
the technical difficulties associated with vortices in superconductors.
In Sec.~\ref{sec:AB}, we review some results from the theory of the Aharonov-Bohm 
effect~\cite{REF:AB} from a point of view due originally to Schulman~\cite{REF:Schulman} in which 
one considers path integrals in multiply-connected spaces. 
In Sec.~\ref{sec:static}, we revisit the BdG eigenproblem from the point of view of gauge transformations
in multiply-connected spaces and isolate the effect of the quasiparticle branch cuts on Feynman paths. 
In Sec.~\ref{sec:branch}, these branch cuts are represented via a functional 
integral over an auxiliary field ${\bf a}$; the theory is constructed to be explicitly invariant under
gauge-transformations associated with this field.  
In Sec.~\ref{sec:dynamic} we extend the results of Sec.~\ref{sec:static} and Sec.~\ref{sec:branch}
to the case of dynamic vortex excitations, finally arriving at an expression for the \BdG\ propagator
(in the presence of vortices) as a gauge-invariant Green function.  
In  Sec.~\ref{sec:discussion} we conclude with a brief discussion of our results.

%\noindent
%{\sl Bogoliubov-De Gennes Equation\/}:

\section{Bogoliubov-De Gennes Equation\/}
\label{sec:bdg}
The QED$_3$ scenario of the pseudogap regime, like the nodal liquid 
scenario~\cite{REF:Balents} which preceded it, focuses on the effect of  
vortex excitations on the quasiparticles of a d-wave superconductor.  In the present
section, we provide motivation by examining the problem of static vortex excitations
in a superconductor.
 The quasiparticle excitations of a superconductor are described by the \BdG\ (BdG) equation
\begin{eqnarray}
\label{eq:bdg}
&&H\pmatrix{u_n \cr v_n}
=E_n\pmatrix{u_n\cr v_n}
\\
&&H\equiv \pmatrix{-(\bm{\nabla}-ie{\bf A})^2-\mu&
{\rm e}^{i\phase/2} \hat{\Delta}{\rm e}^{i\phase/2} \cr
   {\rm e}^{-i\phase/2} \hat{\Delta}{\rm e}^{-i\phase/2}
     & (\bm{\nabla}+ie{\bf A})^2 +\mu},
\label{eq:bdgham}
\end{eqnarray}
where $\mu$ is the chemical potential, $\phase$ is the local superconducting phase,
$e$ is the electronic charge, 
and $\hat{\Delta}$ is the usual d-wave 
pairing operator, which we take to be given by 
$\hat{\Delta} =\Delta_0 \hat{p}_x \hat{p}_y$.  
We have chosen units in which $\hbar^2/2m = 1$, with $m$ being the quasiparticle mass. 
Here, $u_n$ and $v_n$ are  the electron
and hole parts of the BdG wavefunction, respectively.  The electromagnetic gauge
field ${\bf A}$ (which we shall often suppress) and $\phase$ conspire to give this 
theory the following local $U(1)$ symmetry: 
\begin{eqnarray}
A_\mu &\rightarrow&  A_\mu-\frac{1}{e}\partial_\mu \chi,
\\
u_n &\rightarrow& {\rm e}^{i\chi}u_n,
\\
v_n &\rightarrow& {\rm e}^{-i\chi}v_n,
\\
\phase&\rightarrow& \phase+2\chi.
\end{eqnarray}

 In the presence of vortices, $\phase$ has singularities at the locations ${\bf r}_i$ of
each vortex (of vorticity $q_i=\pm1$):
\begin{equation}
\bm{\nabla}\times \bm{\nabla}
\phase ({\bf r})= \sum_i 2\pi q_i \delta^{(2)}({\bf r} -{\bf r}_i).
\label{eq:vortexpos}
\end{equation}
A natural way to proceed with solving  Eq.~(\ref{eq:bdg}) is to attempt
to include the effects of vortex excitations perturbatively by expanding in small
phase gradients.  As $\phase$ appears only in the exponential of the off-diagonal terms, 
one is motivated to perform the following naive gauge transformation
\begin{eqnarray}
\label{eq:naive}
u_n &\to& \tilde{u}_n\equiv{\rm e}^{-i\phase/2}\,u_n,  \\
v_n &\to& \tilde{v}_n\equiv{\rm e}^{i\phase/2}\,v_n. \nonumber
\end{eqnarray}
This effectively moves the phase from the off-diagonal components and introduces it as
a phase gradient in the diagonal components of the BdG equation:
\begin{eqnarray}
\label{eq:bdg2}
&&H'\pmatrix{\tilde{u}_n \cr \tilde{v}_n}
=E_n\pmatrix{\tilde{u}_n\cr \tilde{v}_n},
\\
&&H'\equiv\pmatrix{-(\bm{\nabla}+\frac{i}{2}\bm{\nabla}\phase)^2-\mu&
 \hat{\Delta} \cr
 \hat{\Delta}
     & (\bm{\nabla}-\frac{i}{2}\bm{\nabla}\phase)^2+\mu}.\label{eq:hprimedef}
\end{eqnarray}
However, as discussed by Balents et al~\cite{REF:Balents}, such a gauge transformation
produces branch cuts in the \BdG\ eigenstates.  This can be seen by noting that, if we 
assume that $u_n$ and  $v_n$ are single-valued functions then 
Eq.~(\ref{eq:bdg2}) must be solved under the condition 
that $\tilde{u}_n$ and $\tilde{v}_n$  each gain a factor of ${\rm e}^{i\pi}$ upon encircling
every vortex.  The authors of Ref.~\refcite{REF:Balents} argue that such branch cuts lead to 
frustration effects that would strongly favor the pairing of vortices.  Thus, by only allowing
doubly quantized vortices, there are no branch cuts in the fields $\tilde{u}_n$ and $\tilde{v}_n$ to worry 
about. 
Here, we follow Franz and Tesanovic~\cite{REF:Franz} in assuming that the correct approach is to 
consider the condensation of singly-quantized vortices in a d-wave superconductor.  They avoid 
the branch-cut difficulty by  making a gauge transformation which does not directly introduce branch 
cuts but still keeps track of their physical effects.  Briefly, their technique~\cite{REF:Franz2}
involves splitting the vorticies into two groups
labelled \lq\lq A\rq\rq and \lq\lq B\rq\rq and then performing a gauge
transformation of the form 
\begin{eqnarray}
\label{eq:abgauge}
u_n &\to& u_n{\rm e}^{i\phase_A} \\
v_n &\to& v_n{\rm e}^{-i\phase_B},
\end{eqnarray}
where $\phase_{A(B)}$ is the phase associated only with the vortices in group $A$($B$). 
This gauge transformation has the advantage of not introducing any branch cuts in the 
quasiparticle wave function while at the same time treating the electrons and holes on
an equal footing.  The Berry gauge field ${\bf a}$ emerges, upon averaging over all vortex configurations,
as the difference of phase gradients $\bm{\nabla}\phase_A -\bm{\nabla}\phase_B $.

%-----------------------------------------------------

\section{Path-integral treatment of the Aharonov-Bohm effect\/}
\label{sec:AB}
In Sec.~\ref{sec:bdg} we discussed one difficulty associated with solving
the BdG equation in the presence of vortex excitations, i.e., that naive gauge 
transformations of the form of Eq.~(\ref{eq:naive}) introduce branch cuts in the 
quasiparticle wavefunctions.  To motivate the path integral technique which we shall 
use to handle these branch cuts, in the present section we review previously known 
results~\cite{REF:Schulman,REF:Gerry,REF:Arovas,REF:Kleinert,REF:Kroger} in a related system 
in which gauge transformations must be made with care:
The Aharonov-Bohm problem~\cite{REF:AB}.
Our aim is to find an expression for the propagator $G({\bf x},{\bf y},t)$ 
for an electron constrained to a ring; we represent this physical space by the symbol $M$.  
A thin solenoid penetrates 
the origin and is described by a vector potential
 $A_r = 0$ and $A_{\theta}= \phi/2\pi r$, with $\phi$ being the total flux.  
The Schr\"odinger equation for $G({\bf x},{\bf y},t)$ 
takes the form 
\begin{equation}
\label{eq:gfshulman}
(-i\partial_t - (\bm{\nabla}-ie{\bf A})^2) G({\bf x},{\bf y},t)
=\delta({\bf x}-{\bf y})\delta(t).
\end{equation}
From Feynman's path-integral representation of quantum mechanics, we 
know that $G({\bf x},{\bf y},t)$ may be expressed in terms of a sum
over all paths of an amplitude for each path.  In the present
situation, the doubly-connected nature of $M$ indicates that 
such paths may be divided into homotopically distinct classes depending
on how many times the path in a particular homotopy class winds around the origin.  Thus, 
following Schulman~\cite{REF:Schulman}, we may 
express $G$ in the form 
\begin{equation}
\label{eq:topexp}
G({\bf x},{\bf y},t)=\sum_n G_n({\bf x}_n,{\bf y},t),
\end{equation}
 where the function $G_n({\bf x}_n,{\bf y},t)$ contains only Feynman paths
that wind around the origin $n$ times.  The subscript $n$ on ${\bf x}$ reminds us that
${\bf x}_n$ is the same as ${\bf x}$ after having wound $n$ times around the origin.
Let us consider the physical meaning of 
$G_n({\bf x}_n,{\bf y},t)$.  The space $M$ on which Eq.~(\ref{eq:gfshulman}) is to be solved has 
the topology of a torus.  However, by keeping only Feynman paths which wind $n$ times, it is as if we are 
solving the same equation on a helix-shaped space which winds $n$ times: A helix (topologically, a line) is the 
universal covering space of a torus.  We shall denote the universal covering space by the symbol 
$M^*$; the important properties of $M^{*}$  are as follows: 1) $M^{*}$ is locally 
equivalent to $M$, and 2)  $M^{*}$ is simply connected.  It is further true that the 
Schr\"odinger equation [i.e.~Eq.~(\ref{eq:gfshulman})] is satisfied for each of the terms
$G_n({\bf x}_n,{\bf y},t)$ entering Eq.~(\ref{eq:topexp})~\cite{REF:Laidlaw}.

The fact that  $M^{*}$ is simply connected means that the 
quantity $\exp\left( i e\int_{{\bf y}}^{{\bf x}_n} {\bf A}\cdot d{\bf s}\right)$ is 
well-defined on it and can be used to simplify the equation for 
$G_n$.  Thus, by writing 
\begin{equation}
G_n({\bf x}_n,{\bf y},t) 
=\bar{G}_n({\bf x}_n,{\bf y},t)
\exp\left( i e\int_{{\bf y}}^{{\bf x}_n} {\bf A}\cdot d{\bf s}\right), 
\end{equation}
it can be seen that 
$\bar{G}_n$ satisfies 
\begin{equation}
\label{eq:gfshulman2}
(-i\partial_t - \bm{\nabla}^2) \bar{G}_{n}({\bf x}_n,{\bf y},t)
=\delta({\bf x}_n-{\bf y})\delta(t),
\end{equation}
i.e., it is the Green function for the free-particle (by this we mean ${\bf A}={\bf 0}$) Schr\"odinger
equation on $M^{*}$, which we denote by $\bar{G}^{(0)}_{n}({\bf x}_n,{\bf y},t)$
and has the explicit form~\cite{REF:Gerry,REF:Arovas,REF:Kleinert,REF:Kroger} 
\begin{equation}
\label{eq:freepartprop}
\bar{G}^{(0)}_{n}({\bf x}_n,{\bf y},t) 
= \frac{1}{4\pi t}{\rm e}^{\frac{i}{4t}(x^2 + y^2)}
\int_{-\infty}^{\infty} d\lambda \,{\rm e}^{i\lambda(\theta'-\theta+2\pi n) }
I_{|\lambda|}\left(\frac{xy}{2it}\right),
\end{equation}
where $I_{\lambda}$ is the modified Bessel function and 
we have written the coordinates ${\bf x}_n$ and ${\bf y}$ in terms of the radial 
coordinates $(x,\theta'+2\pi n)$ and  $(y,\theta)$, respectively.  We emphasize that this 
is not the usual two-dimensional free particle propagator, but that part of the propagator 
due to Feynman paths with winding number $n$ (or, equivalently, the free propagator 
on $M^*$).  The usual free-particle propagator may be obtained from 
Eq.~(\ref{eq:freepartprop}) by summing over all winding numbers $n$. 

The next step is to insert $\bar{G}^{(0)}_n({\bf x}_n,{\bf y},t)$
into Eq.~(\ref{eq:topexp}) after having directly calculated
the factor $\exp\left( i e\int_{{\bf y}}^{{\bf x}_n} {\bf A}\cdot d{\bf s}\right)$:
\begin{eqnarray}
\label{eq:topexp2a}
\!\!\!\!G({\bf x},{\bf y},t)\!&=&
\sum_n {\rm e}^{ie\phi(\theta'-\theta+2\pi n)/2\pi} \bar{G}^{(0)}_n({\bf x}_n,{\bf y},t)
\\
&=&\!\!\!\frac{1}{4\pi t}{\rm e}^{\frac{i}{4t}(x^2 + y^2)} \!
\sum_m {\rm e}^{im(\theta'-\theta)}\!
I_{|m-\alpha|}\!\left(\frac{xy}{2it}\right),
\label{eq:topexp2b}
\end{eqnarray}
where $\alpha \equiv e\phi/2\pi$ measures the flux through the hole.  
Equation~(\ref{eq:topexp2b}) is the single-particle propagator for the Aharonov-Bohm
problem~\cite{REF:Gerry,REF:Arovas,REF:Kroger}, which is obtained from 
Eq.~(\ref{eq:topexp2a}) using Eq.~(\ref{eq:freepartprop}) along with the Poisson 
summation formula.

For our purposes, we are primarily interested in 
Eq.~(\ref{eq:topexp2a}), which exhibits the structure proposed by Schulman~\cite{REF:Schulman}: The
propagator is expressed as a sum of terms, each of which is a 
{\it free\/} propagator associated with a particular winding number (i.e.~on the covering space)
multiplied by a gauge-field dependent factor.  
Before proceeding, however, let us briefly note one feature
of Eq.~(\ref{eq:topexp2b}), namely that for integral $\alpha $, the summation over $m$ may be shifted by $\alpha$
and  Eq.~(\ref{eq:topexp2b}) becomes the two-dimensional free propagator, i.e., the solution to 
Eq.~(\ref{eq:gfshulman2}) in free space.  This is merely the statement that for integral $\alpha$ there
are no interference effects associated with the flux and the introduction
of the universal covering space was unneccesary.  For non-integral $\alpha$, however, 
the universal covering space provides a natural way to identify Feynman paths in different homotopy 
classes which  have different phase factors arising from the gauge field.  
In a similar fashion, we shall see that our technique for handling vortex excitations in d-wave superconductors 
by invoking the covering space is useful for the case of $hc/2e$ vortices [for which the gauge
transformation in Eq.~(\ref{eq:naive}) produced branch cuts]  but unnecessary for doubly quantized
vortices [for which the gauge
transformation in Eq.~(\ref{eq:naive}) produced no branch cuts].
In the next section, we shall exploit the universal covering space to gain insight into the problem of 
$hc/2e$ vortices in
d-wave superconductors.

%\noindent
%{\sl Vortices in a d-wave superconductor\/}:

\section{Vortices in a d-wave superconductor: Static case\/}
\label{sec:static}
Having discussed the covering-space approach to the Aharonov-Bohm problem, in the present section we apply these 
ideas to the context of multiple vortices in a d-wave superconductor.   In this section we shall 
assume a high-temperature approximation in which the vortices can be taken to be static. The extension
to quantum-vortex fluctuations will be straightforward and discussed in Sec.~\ref{sec:dynamic}.
Our starting point is the following expression for the BdG propagator averaged over vortex 
positions:
\begin{eqnarray}
\label{eq:funcintstat}
&&[G({\bf r},{\bf r}';\tau) ]_{\phase} \equiv \int \curD \phase\,
 G({\bf r},{\bf r}';\tau) {\rm e}^{-S[\phase]},
\\
\label{eq:green}
&& (\partial_{\tau}+H) G({\bf r},{\bf r}';\tau) 
= \delta^{(2)}({\bf r}-{\bf r}') \delta(\tau),
\end{eqnarray}
where $G({\bf r},{\bf r}';\tau)$ is the BdG Green function in 
the presence of a fixed pattern of vortices and  $S[\phase]$ is the Boltzmann weight associated with a 
particular, static, pattern of vortices.  The Hamiltonian $H$ is given by Eq.~(\ref{eq:bdgham}); however, henceforth
we shall not display the physical gauge field ${\bf A}$. It may be reinstated at the end of the calculation.

Each vortex configuration entering into Eq.~(\ref{eq:funcintstat}) corresponds to a particular set of locations
for the ${\bf r}_i$ in Eq.~(\ref{eq:vortexpos}).  Thus, as in the Aharonov-Bohm problem, the manifold
$M$ on which we aim to solve Eq.~(\ref{eq:green}) is multiply connected, i.e., it has \lq \lq holes\rq\rq\ at each 
${\bf r}_i$.  To ascertain the physical importance of this multiple-connectedness, we exchange 
 the original multiply-connected space $M$ (on which Eq.~(\ref{eq:green}) is defined)
for its simply-connected covering space $M^{*}$ 
on which a gauge transformation of the form of Eq.~(\ref{eq:naive}) may be performed.
 As we saw in Sec.~\ref{sec:bdg},
for the Aharonov-Bohm effect the relevant universal covering space was 
a line.  
In the present case, the fact that we shall always consider the case of multiple vortices 
means that the associated universal covering space is difficult to envision physically.   
Fortunately, as in the case of
the Aharonov-Bohm effect, the covering space will merely serve as a 
way to keep track of the phase picked up by various Feynman 
paths.  

Following our procedure in the preceding section, we express 
$G({\bf r},{\bf r}';\tau)$ as a sum of propagators, the Feynman 
paths of which are grouped into topologically distinct sectors:
\begin{equation}
\label{eq:sectors}
G({\bf r},{\bf r}',\tau)=
\sum_{\bf n} G_{\bf n}({\bf r}_{\bf n},{\bf r}';\tau),
\end{equation}
where the vector index ${\bf n}=(n_1,n_2,\cdots) $ keeps 
track of the winding number $n_i$ associated with the $i$th vortex.
Technically speaking, this winding number does not fully label the elements of the non-abelian
homotopy group associated with the multiply-connected space in question, but this does not 
matter for the calculation we are attempting since all that matters is that there exists a label
for these elements. 
On $M^{*}$ we have at our disposal the quantity 
\begin{equation} 
\label{eq:disposal}
\chi_{\bf n}({\bf r}_{\bf n}) \equiv \int_{\bf 0}^{{\bf r}_{\bf n}} 
\bm{\nabla}\phase(\bar{\bf r})   \cdot d\bar{\bf r},
\end{equation}
which we emphasize is not uniquely defined on the space $M$. 
For simplicity, we have set ${\bf r}'={\bf 0}$.
 Physically,
$\chi_{\bf n}$ represents a line integral along a particular path
through the vortices.  We write 
$G_{\bf n}({\bf r}_{\bf n},{\bf 0},\tau)$ in the form
\begin{equation}
\label{eq:gaugetrans1}
G_{\bf n}({\bf r}_{\bf n},{\bf 0},\tau)= 
\exp\left[\frac{i}{2}\chi_{\bf n}({\bf r}_{\bf n}) \hat{\sigma}_3\right]
\tilde{G}_{\bf n}({\bf r}_{\bf n},{\bf 0},\tau),
\end{equation}
where $\hat{\sigma}_3$ is the usual Pauli matrix.
As in Sec.~\ref{sec:AB}, this amounts to making  different gauge transformations for Feynman 
paths having different winding numbers.  The 
$\tilde{G}_{\bf n}$ satisfy 
\begin{equation}
\left(
\partial_{\tau}+
H'
 \right)
\tilde{G}_{\bf n}({\bf r}_{\bf n},{\bf 0};\tau) = \delta^{(2)}({\bf r}_{\bf n})\delta(\tau),
\end{equation}
where $H'$ is given by Eq.~(\ref{eq:hprimedef}), i.e., the Hamiltonian obtained via the naive gauge transformation of 
Sec.~\ref{sec:bdg}.

%------------------------------
\begin{figure}[hbt]
% \vskip0.50cm
 \epsfxsize=2.5in 
\epsfxsize=2.9in
%  \epsfxsize=\columnwidth
%\centerline{\epsfbox{FIGS/fig1.eps}}
\centerline{\epsfbox{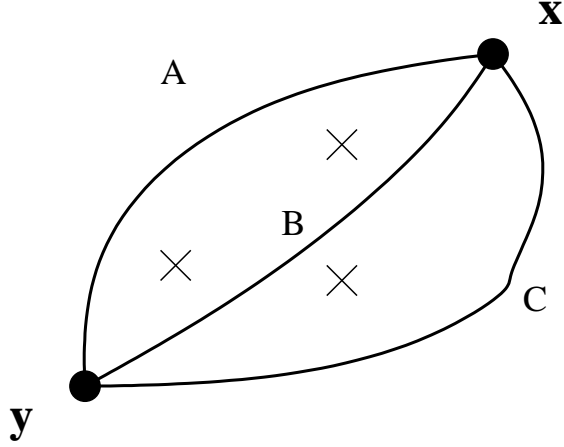}}
\vskip0.50cm
\caption{
Sketch of various homotopically inequivalent Feynman paths (black curves) connecting 
the points ${\bf y}$ and ${\bf x}$ in the presence of 
vortices (represented by large crosses). As an example of how the winding number $n$ is defined, we note that
for the leftmost vortex $n$ can be taken to be $0$ for path A and $1$ for path B.
  }  
\label{FIG:paths}
\end{figure}
%------------------------------

Having made a gauge transformation on the covering space, we proceed
by computing the path-dependent gauge-transformation factor
$\exp\left[\frac{i}{2}\chi_{\bf n}({\bf r}_{\bf n}) \hat{\sigma}_3\right]$.  
The path-dependence of $\chi_{\bf n}({\bf r}_{\bf n})$ 
is exhibited schematically in Fig.~\ref{FIG:paths}.  Consider for example the loop made by 
paths B and C: since there is one vortex inside the loop, we have 
\begin{equation}
\int_{{\rm B} - {\rm C}} \bm{\nabla}\phase(\bar{\bf r} ) \cdot d\bar{\bf r} = \pm 2\pi ,  
\end{equation}
depending on the vorticity of the enclosed vortex.  Thus, $\chi_{\bf n}({\bf r}_{\bf n})$ 
differs by $2\pi$ for paths B and C.  
More generally, we may define
\begin{eqnarray}
\chi_{\bf n}({\bf r}_{\bf n})& =& 
\phase({\bf r}) - \phase({\bf 0}) + 2\pi \sum_i \, q_i\,n_i,
\\
\label{eq:step}
{\rm e}^{\frac{i}{2}\chi_{\bf n}({\bf r}_{\bf n}) \hat{\sigma}_3}
&=& {\rm e}^{\frac{i}{2}(\phase({\bf r}) - \phase({\bf 0}))\hat{\sigma}_3}
 {\rm e}^{i \pi \sum_i \, q_i n_i},
\end{eqnarray}
where it is important to note that in Eq.~(\ref{eq:step}) we have used the fact that 
$\exp( i \pi n\hat{\sigma}_3)=\exp( i \pi n)$ for integer $n$.
We thus have the following expression for the averaged BdG Green function:
 \begin{equation}
\label{eq:bdgint}
[G({\bf r},{\bf r}';\tau) ]_{\phase} =
 \int \curD \phase\,{\rm e}^{-S[\phase]}{\rm e}^{\frac{i}{2}\phase({\bf r})\hat{\sigma}_3}
\left[\sum_{\bf n} 
 \tilde{G}_n({\bf r}_{\bf n},{\bf r}';\tau)
{\rm e}^{i \pi \sum_i \,q_i n_i} \right]
{\rm e}^{-\frac{i}{2}\phase({\bf r}')\hat{\sigma}_3}.
\end{equation}
Since we are considering 
only singly quantized vortices, so that $q_i = \pm 1$, the  
branch-cut factor  $\exp i \pi \sum_i \, q_i n_i $ is given by $\pm 1$ for different trajectories and is the
manifestation, within the present approach, of the quasiparticle branch cuts encountered 
in Sec.~\ref{sec:bdg}.  In Fig.~\ref{FIG:paths}, since they are separated by {\it two} vortices, 
the branch-cut factor is the same for paths A and B.  In general, 
if we considered the condensation of {\it pairs} of vortices ($q_i = \pm 2$),
then these factors could be all taken to be unity and there would be no need to invoke 
the covering space construction.  However, although there is a $Z_2$ character to the physics of fluctuating
branch cuts as discussed here, it has a origin which is distinct from that of
the $Z_2$ gauge theory of Senthil and Fisher~\cite{REF:Senthil}, which, like the theory of
 Ref.~\refcite{REF:Balents}, only
incorporates doubly quantized vortices.

Before proceeding to the next stage of the calculation, we pause to note that 
the argument of the functional integral in Eq.~(\ref{eq:bdgint}) 
is the analogue, within the present context, of Eq.~(\ref{eq:topexp2a}) in the calculation
of the Aharonov-Bohm propagator in Sec.~\ref{sec:AB}.  For that case it is possible to calculate
the propagator on the covering space [cf.~Eq.~(\ref{eq:freepartprop})] explicitly.
In the present case, however, $\tilde{G}_n({\bf r}_{\bf n},{\bf r}';\tau)$ is most likely not analytically 
solvable and, in the next section, we proceed by making some simplifying approximations.

\section{Effect of branch cuts on quasiparticle dynamics}
\label{sec:branch}

In the present section, we shall attempt to evaluate the functional integral over vortex and spin-wave 
like phase fluctuations in Eq.~(\ref{eq:bdgint}).
As discussed in Sec.~\ref{sec:static},
the effect of the branch cuts in the quasiparticle wavefunctions
has been traced back to the \lq\lq branch-cut\rq\rq\ factor  $\exp i \pi \sum_i \,q_i n_i$ 
in Eq.~(\ref{eq:step}) which differs for homotopically inequivalent trajectories.  
If this factor were absent (e.g.~if we had only allowed the proliferation of 
doubly-quantized vortices 
so that, in effect, $\sum_i \,q_i n_i$ would be constrained to be an even integer), then there
would be no need to divide the equation for the propagator
into topologically inequivalent sectors, and the gauge transformation given in 
Eq.~(\ref{eq:naive}) could have been made on $M$.

Let us turn to the evaluation
of the functional integral over phase fluctuations.  Formally, $\varphi$ appears in two distinct
places in Eq.~(\ref{eq:bdgint}): 
Firstly, it appears in the definition of $H'$ in Eq.~(\ref{eq:hprimedef}).  As our aim is to focus on 
the effect of branch cuts, here we neglect $\bm{\nabla}\phase$ in $H'$.  This simplifying approximation 
is not technically necessary at this point.  It is motivated by the fact that, were we to keep 
 the local phase gradient as another (\lq\lq Doppler\rq\rq) gauge field $\bf v$ then, as
discussed in Refs.~\refcite{REF:Franz,REF:Herbut}, $\bf v$ would end up being irrelevant in 
 the renormalization group sense.
Secondly, $\varphi$ appears in the exponential
factors $\exp\left(\pm i\phase({\bf r})\hat{\sigma}_3/2\right)$ in Eq.~(\ref{eq:bdgint}).  
Let us examine the 
matrix structure of 
the argument of the functional integral by writing it as a $2\times 2$ matrix:
 \begin{equation}
{\rm e}^{\frac{i}{2}\phase({\bf r})\hat{\sigma}_3}
\sum_{\bf n} 
 \tilde{G}_n
 {\rm e}^{-\frac{i}{2}\phase({\bf r}')\hat{\sigma}_3}
=\pmatrix{\tilde{G}_{n,11}{\rm e}^{\frac{i}{2}(\phase({\bf r})-\phase({\bf r}'))}  
&\tilde{G}_{n,12} {\rm e}^{\frac{i}{2}(\phase({\bf r})+\phase({\bf r}'))} \cr
\tilde{G}_{n,21} {\rm e}^{-\frac{i}{2}(\phase({\bf r})+\phase({\bf r}'))} 
&\tilde{G}_{n,22} {\rm e}^{-\frac{i}{2}(\phase({\bf r})-\phase({\bf r}'))} }.
\end{equation}
By writing out the full matrix form for the product of these three factors appearing in Eq.~(\ref{eq:bdgint})
we see that whereas the diagonal components contain a phase difference, the off-diagonal components contain a 
sum of phases and thus must vanish upon performing the average over the $\phase$.  This merely implies 
that the off-diagonal components of the single-particle Green function must vanish in the pseudogap regime, i.e.,
that superconductivity has been destroyed.  
Being phase differences, the phase factors appearing in the diagonal components do not vanish upon 
averaging over ${\phase}$.  We shall take them to be unity, invoking the same argument which we used for the 
phase gradients in $H'$.

By making the preceding approximations, we are  assuming that the branch cuts 
factors lead to quantum interference effects that 
 have a dominant effect on the low-energy quasiparticles.  We thus arrive at 
 \begin{eqnarray}
\nonumber
&&[G({\bf r},{\bf r}';\tau) ]_{\phase} \simeq
 \int \curD \phase\,{\rm e}^{-S[\phase]}\sum_{\bf n} {\rm e}^{i \pi \sum_i \,q_i  n_i}
\\
 \label{eq:bdgint2}
&& \times\left.
\langle {\bf r}_n,\tau | \left[\partial_{\tau} + \pmatrix{-\nabla^2-\mu&
\Delta_0p_x\,p_y  \cr
\Delta_0   p_x\,p_y
     & \nabla^2+\mu}\right]^{-1} |{\bf r}',0 \rangle \right|_{\rm diag.},
\end{eqnarray}
where we have explicitly written the expression for $\tilde{G}_n({\bf r}_{\bf n},{\bf r}';\tau)$
within this approximation.  The subscript \lq\lq diag.\rq\rq\ indicates that our expression 
for $[G({\bf r},{\bf r}';\tau) ]_{\phase}$ only includes the diagonal components of the 
expression on the right side of Eq.~(\ref{eq:bdgint2}), the off-diagonal components being zero
as noted above.  Henceforth, we shall suppress this subscript, although all subsequent equations share this
property that only the diagonal components are of physical interest.

%-------------------------

The next step is to represent the branch-cut factor in terms of the flux of a field ${\bf a}$, allowing 
us to evaluate the functional integral over vortex configurations in Eq.~(\ref{eq:bdgint2}).
 To do this, we define ${\bf a}$ such that its local \lq\lq magnetic field\rq\rq\ 
has singularities at the locations of all the vortices (but curl-free on $M^*$):
\begin{equation}
\bm{\nabla}\times
{\bf a} ({\bf r})= \pi \sum_i q_i \delta^{(2)}({\bf r} -{\bf r}_i).
\end{equation}
To account for the factor ${\rm e}^{i  \pi \sum_i \,q_i n_i}$, we write 
$\sum_i \, q_i n_i$ as the flux through a carefully chosen loop:
\begin{equation}
\label{eq:split1}
 \sum_i \, q_i n_i = -\frac{1}{\pi} \int_{\bf n} {\bf a} \cdot d{\bf r} +
                 \frac{1}{\pi} \int_{\Gamma} {\bf a} \cdot d{\bf r},
\end{equation}
where the subscript ${\bf n}$ indicates that the first integral winds
along the path falling in the homotopy class labelled by ${\bf n}$
(i.e.~from $\bf y$ to ${\bf x}_n$ in $M^*$), and
the subscript $\Gamma$ indicates that the integral is to be taken along
some arbitrary fixed reference path ${\bf x}_{\Gamma}(s)$:
\begin{equation}
\int_{\Gamma} {\bf a} \cdot d{\bf r} \equiv \int_0^1 ds a_{\mu}({\bf x}_{\Gamma}(s)) \frac{x_{\Gamma,\mu}(s)}{ds},
\label{eq:gammadef}
\end{equation}
where ${\bf x}_{\Gamma}(0)={\bf r}'$ and ${\bf x}_{\Gamma}(1)={\bf r}$.
Thus, we have 
\begin{eqnarray}
&&
 \nonumber
[G({\bf r},{\bf r}';\tau) ]_{\phase} \simeq
 \int \curD \phase\,{\rm e}^{-S[\phase]}
 \int \curD {\bf a}     \,{\rm e}^{i\int_{\Gamma} {\bf a} \cdot d{\bf r}}
\sum_{\bf n}  {\rm e}^{-i  \int_{\bf n} {\bf a} \cdot d{\bf r}}
\\
\nonumber
&&\qquad \times
\langle {\bf r}_n,\tau | \left[\partial_{\tau} + \pmatrix{-\nabla^2-\mu&
\Delta_0 p_x\,p_y  \cr
   \Delta_0 p_x\,p_y
    & \nabla^2+\mu}\right]^{-1} |{\bf r}',0 \rangle 
 \\
&&\qquad\times
\delta(\bm{\nabla}\times{\bf a} ({\bf r})- \pi \sum_i q_i \delta^{(2)}({\bf r} -{\bf r}_i)),
\label{eq:sectors2}
\end{eqnarray}
where we must emphasize that the field ${\bf a}$ has entered in a  gauge-invariant 
way, in the sense that $[G({\bf r},{\bf r}';\tau) ]_{\phase}$ is manifestly dependent
only on ${\bf \nabla}\times {\bf a}$.
Although the quantity $\int_{\bf n} {\bf a} \cdot d{\bf r}$
in Eq.~(\ref{eq:split1}) is trajectory-dependent,
 the integrals along the path $\Gamma$ are by definition 
the same for each term in Eq.~(\ref{eq:sectors2}).  Before proceeding, we remark that this propagator
is reminiscent of the gauge-invariant Green function discussed in Ref.~(\refcite{REF:Altshuler}).

Next, we make another
gauge transformation of the form of Eq.~(\ref{eq:gaugetrans1}), absorbing the
factors $\exp -i  \int_{\bf n} {\bf a} \cdot d{\bf r}$
into the Green function associated with that 
particular trajectory (or, rather, class of trajectories associated with a particular
homotopy class in the presence of the vortex excitations).  After making such a gauge 
transformation, the division of propagators into topologically distinct sectors is
redundant (the field ${\bf a}$ keeps track of the branch cut factors),
leaving us with the following expression for the Green function:
\begin{eqnarray}
\nonumber 
&&[G({\bf r},{\bf r}';\tau) ]_{\phase} \simeq 
\int \curD \phase\,{\rm e}^{-S[\phase]}
 \int \curD {\bf a}     \, {\rm e}^{i\int_{\Gamma} {\bf a} \cdot d{\bf r}}
 \langle {\bf r},\tau 
 | \frac{1}{
\partial_{\tau} + 
H[{\bf a}]}
 |{\bf r}',0 \rangle 
\label{eq:hofa}
\\
&&\qquad\qquad\times
\delta(\bm{\nabla}\times{\bf a} ({\bf r})- \pi \sum_i q_i \delta^{(2)}({\bf r} -{\bf r}_i)),
\label{eq:sectors3}
\\
&&H[{\bf a}]\equiv
\pmatrix{({\bf p}+{\bf a})^2-\mu&
\hat{D}  \cr
  \hat{D}
     & -({\bf p}+{\bf a})^2+\mu},
\\
&& \hat{D} \equiv \frac{\Delta_0}{2} [(p_x + a_x)(p_y + a_y)\!
+\!(p_y + a_y)(p_x+ a_x)] .
%\nonumber
\end{eqnarray}
%---------------------
The most straighforward way to verify the equality of Eq.~(\ref{eq:sectors2}) and Eq.~(\ref{eq:sectors3}) is 
to examine the gauge-transformation properties of the arguments of the respective functional integrals.
  The next step is to evaluate the functional integral
over vortex and spin-wave like excitations, leaving an effective functional integral over 
the field ${\bf a}$.  Formally, we may follow  Franz and collaborators~\cite{REF:Franz},
writing 
\begin{equation}
\label{eq:actionberry}
{\rm e}^{-S_{\bf a}} \equiv  \int \curD \phase \, {\rm e}^{-S[\phase]}
\delta(\bm{\nabla}\times{\bf a} ({\bf r}) -  
2\pi \sum_i q_i \delta^{(2)}({\bf r} -{\bf r}_i)).
\end{equation}
%-----------------------------------
Using this definition of the vortex action, we arrive at our final expression for the 
single-particle BdG propagator in the presence of static vortex excitations:
\begin{equation}
\label{eq:finalstatic}
 [G({\bf r},{\bf r}';\tau) ]_{\phase} \simeq
 \int \curD {\bf a}     \,{\rm e}^{-S_{\bf a}}\,
  {\rm e}^{i\int_{\Gamma} {\bf a} \cdot d{\bf r}}
 \langle {\bf r},\tau | 
\frac{1}{
\partial_{\tau} + 
H[{\bf a}]}
 |{\bf r}',0 \rangle .
\label{eq:greenfinal2}
\end{equation}
Equation~(\ref{eq:finalstatic}) is essentially our main result, i.e., we have obtained a 
gauge-theory model for vortex excitations in a d-wave superconductor via a path-integral technique.
However, we recall that until now we have considered only static vortex
excitations, so that the functional integral over vortex excitations in, e.g., 
Eq.~(\ref{eq:funcintstat}), is truly a thermal average.  As we shall see, however, 
the generalization to arbitrary dynamic vortex excitations is straighforward within the 
path-integral technique.  In the next section, we discuss this issue in detail.

\section{Vortices in a d-wave superconductor: Dynamic case\/}
\label{sec:dynamic}
In arriving at  Eq.~(\ref{eq:finalstatic}), we have made use of the 
Feynman path integral description of the BdG propagator in the presence 
of a static pattern of vortices.  In the present section, we generalize this
procedure to the case of arbitrary fluctuating vortices.  The quantity we are interested 
in calculating is, formally, the same as was discussed in
 Secs.~\ref{sec:static} and \ref{sec:branch} and is given by
\begin{eqnarray}
[G({\bf r},{\bf r}';\tau) ]_{\phase} &\equiv& \int \curD \phase \,{\rm e}^{-S[\phase]}
 \langle {\bf r},\tau |
\frac{1}{
\partial_{\tau} + 
H }
| {\bf r}',0\rangle, 
\label{eq:dynamicgreen}
\\
H &\equiv&
\pmatrix{-\nabla^2-\mu&
{\rm e}^{i\phase/2} \hat{\Delta}{\rm e}^{i\phase/2} \cr
   {\rm e}^{-i\phase/2} \hat{\Delta}{\rm e}^{-i\phase/2}
     & \nabla^2+\mu},
\end{eqnarray}
where the only distinction from the preceding discussion is that now $\phase$ is  time-dependent and $S[\phase]$ 
is an action, as opposed to a Boltzmann.

In the picture discussed in the preceding sections, the 
various Feynman paths were parametrized by the temporal variable 
and fell into homotopically inequivalent classes defined by the locations
of the (static) vortices.  
Now imagine that the vortices are time-dependent.  One may of course still construct
a conventional Feynman path integral even though $H$ is time-dependent.  
However, since the vortex positions evolve in time, Feynman paths parametrized by
time do not fall into homotopically inequivalent classes in the same way.
Thus, it is convenient to use a path-integral method
which treats the time variable $\tau$ on an equal footing with the spatial coordinates.
This may be realized by using a so-called \lq\lq fifth-parameter\rq\rq\ path integral
scheme~\cite{REF:Schulman}.  Within such a scheme, one introduces a ficticious
temporal coordinate $\lambda$ and constructs a path integral representation of
the corresponding Green function in which the temporal variable in the 
Feyman paths is $\lambda$, not $\tau$.
For example, consider the following Schr\"odinger equation:
\begin{eqnarray}
\label{eq:gammaschr}
&&(i\partial_{\lambda} + \curL)\,\Gamma(x,y;\lambda) = -i\delta(x-y) \delta(\lambda),
\\
&&\curL \equiv \partial_{\tau}+H ,
\end{eqnarray}
where $x \equiv ({\bf r},\tau)$ and $y\equiv ({\bf r}',0)$.  
It is clear that the propagator $\Gamma(x,y;\lambda)$ may be expressed as a path integral in which the
paths are labeled by $\lambda$.  Furthermore, it may be simply related to the argument of 
Eq.~(\ref{eq:dynamicgreen}) via
\begin{equation}
 \langle {\bf r},\tau |
\frac{1}{
\partial_{\tau} + 
H
 }
| {\bf r}'0\rangle = \int_0^{\infty} d\lambda \,\Gamma(x,y;\lambda){\rm e}^{-\epsilon \lambda},
\end{equation}
where $\epsilon = 0+$.

The utility of the preceding discussion is that now we have a path-integral representation
(i.e.~that of Eq.~(\ref{eq:gammaschr})
for the argument of Eq.~(\ref{eq:dynamicgreen}) in which the variable $\tau$ is treated like any 
other coordinate, so that instead of thinking of vortices in a two-dimensional $XY$ model we 
envision vortex loops in a three-dimensional $XY$ model.  The vortices are \lq\lq static\rq\rq\
 relative to the coordinate $\lambda$, allowing us to split the Feynman paths into homotopically
distinct sectors depending on how they wind relative to them.  This amounts to constructing 
an equation for $\Gamma(x,y;\lambda)$ that is of the form of Eq.~(\ref{eq:sectors}).  Following the same 
steps as in the static case (i.e, making gauge transformations on the covering space, etc.), 
generalized from $2$ to $3$ dimensions, we find that 
 $[G({\bf r},{\bf r}';\tau) ]_{\phase}$ is given by
\begin{eqnarray}
\label{eq:dynamic}
&&[G({\bf r},{\bf r}';\tau) ]_{\phase} \simeq
 \int \curD a     \,{\rm e}^{-S_a} {\rm e}^{i\int_{\Gamma} {\bf a} \cdot d{\bf r}}
 \langle {\bf r},\tau | 
\frac{1}{\partial_{\tau} + ia_{\tau}+
H[{\bf a}]}
 |{\bf r}',0 \rangle 
 \\
&&S_a = -K(\partial_{\nu} a_{\mu} - \partial_{\mu} a_{\nu})^2/4,
\label{eq:actionberry2}
\end{eqnarray}
where now the vector field $a$ has three components given by $a = (a_{\tau},a_x,a_y)$.  
The boldface quantity ${\bf a}$ still refers to the spatial vector  ${\bf a} = (a_x,a_y)$,
and $H[{\bf a}]$ is still given by Eq.~(\ref{eq:hofa}) (although we again emphasize that 
now $a$ fluctuates dynamically).  The line integral with subscript label $\Gamma$ is similarly generalized
from Eq.~(\ref{eq:gammadef}) to now indicate a line integral from $({\bf r}',0)$ to   $({\bf r},\tau)$
The gauge-invariant action $S_a$ (with $K$ being an appropriate coupling)
is formally defined by an equation exactly analagous to 
Eq.~(\ref{eq:actionberry}); in Eq.~(\ref{eq:actionberry2}) we have used the results of Franz 
et al.~\cite{REF:Franz} who computed it using (see also Ref.~\refcite{REF:Herbut}) a model of 
free fluctuating vortex loops.  
It is important to note however that this form for $S_a$  
could be expected on general symmetry grounds, it being the simplest quadratic
action which respects gauge invariance.

To make the connection between $[G({\bf r},{\bf r}';\tau) ]_{\phase} $  and
the propagator for the QED$_3$ gauge theory (i.e.~to make the gauge symmetry manifest) 
we represent $[G({\bf r},{\bf r}';\tau) ]_{\phase} $ 
in terms of a fermionic path integral
\begin{eqnarray}
\label{eq:gdef}
&&[G({\bf r},{\bf r}';\tau) ]_{\phase}
 \simeq \int \curD\psi \curD\psi^{\dagger}\curD a  \,
{\rm e}^{-S_{\psi}-S_a}
\psi({\bf r},\tau) \psi^{\dagger}({\bf r}',0) {\rm e}^{i\int_{\Gamma} {\bf a} \cdot d{\bf r}},
\\
&& S_{\psi} \equiv\int d^3x \psi^{\dagger} \left[\partial_{\tau} + ia_{\tau}+
H[a]
\right] \psi ,
\end{eqnarray}
where $\psi^{\dagger}$ and $\psi$ represent Nambu spinor fields.  
The action  $S_{\psi}+ S_a$ is invariant under the gauge 
transformation $\psi \to {\rm e}^{-i\chi}\psi, a_{\mu} \to a_{\mu} +\partial_{\mu}\chi$.  In addition, 
we see that the  factor ${\rm e}^{i\int_{\Gamma} {\bf a} \cdot d{\bf r}}$ in the argument of 
Eq.~(\ref{eq:gdef}) ensures that 
\begin{equation}
\psi({\bf r},\tau) \psi^{\dagger}({\bf r}',0) {\rm e}^{i\int_{\Gamma} {\bf a} \cdot d{\bf r}},
\end{equation}
and therefore $[G({\bf r},{\bf r}';\tau) ]_{\phase}$, is also gauge-invariant.  One may also 
expand the fermions near the nodes of the d-wave order parameter; by combining the Nambu 
fields $\psi$ into four component Dirac spinors~\cite{REF:Franz} one may complete the connection
to QED$_3$.  However, for the present purposes it is sufficient to have the above gauge theory 
be our final expression.

\section{Discussion\/}
\label{sec:discussion}
In this Paper, we have explored the effect of vortices on the
the quasiparticle excitations of d-wave superconductors in an attempt to understand the 
pseudogap phenomena.  
Our final expression for the single-particle Green 
function (i.e.~Eq.~(\ref{eq:dynamic})) makes direct contact with the  QED$_3$~\cite{REF:Franz}
scenario of the pseudogap phenomena.  
As discussed in Sec.~\ref{SEC:intro}, the original formulation of the QED$_3$ of the pseudogap 
phenomena~\cite{REF:Franz} relied on splitting the vortices
into two groups ($A$ and $B$) and making a gauge transformation of the electrons with respect to group
$A$ and the holes with respect to  $B$.  The aim of such an approach is to avoid directly introducing
branch cuts in the quasiparticle wavefunctions.   

The approach presented here deals with the quasiparticle branch cuts in a more direct fashion, by
realizing that the problem of solving the BdG equation in the presence of vortices may be formulated 
as a problem of
making gauge transformations on multiply connected manifolds.  By going to the universal covering space of the 
associated manifold, the branch cuts emerge in a controllable fashion as factors of $\pm 1$ multiplying 
various Feynman paths.  The field ${\bf a}$ arises in an attempt to mimic the effect of 
these branch-cuts; in particular the line integral factor in Eq.~(\ref{eq:gdef}) comes from 
expressing the branch cut factors $\pm 1$  associated with various Feynman
paths in terms of the flux of $a$.
Although we have clarified the origin of the Berry gauge field in the BdG {\it action\/}, we
have not clarified issues associated with the calculation of the single particle Green function 
[i.e.~Eq.~(\ref{eq:gdef})]
 As has been noted in several recent 
papers~\cite{REF:Khveshchenko,REF:Marcelpreprintnote,REF:Kh2}, there are many technical 
difficulties associated with the calculation of this quantity due to the presence of the line integral
over the gauge field.  As noted in a recent preprint~\cite{REF:marcel3}, however, for certain
gauge-invariant quantities these line integrals do not appear.  
%

%---------------------------
\noindent
{\sl Acknowledgments\/}:  We gratefully acknowledge 
stimulating and enjoyable discussions with 
Marcel Franz, Alexandre Zagoskin, \.{I}nan\c{c}~Adagideli, Paul Goldbart, Tamar Pereg-Barnea, Tom Davis, Igor Herbut
and Mark Laidlaw.  This work 
was supported by NSERC.
%---------------------------
%---------------------------
%---------------------------

%------------------------------
%\end{multicols}
%------------------------------
%------------------------------
\end{document}